\documentclass[useAMS]{mn2e}
\ifx\pdfoutput\thisisundefined
   \newcommand{\dofig}{P}
\else
   \newcommand{\dofig}{D}
\fi

\if\dofig F
   \newcommand{\putfig}[2]{\vspace{#2}}
\else
   \if\dofig D
      \usepackage[pdftex]{graphicx}
      \newcommand{\putfig}[2]{\includegraphics[width=#2]{#1.pdf}}
   \else
      \usepackage[dvips]{graphicx}
      \newcommand{\putfig}[2]{\includegraphics[width=#2]{#1.eps}}
   \fi
\fi

\def\farcs{\hbox{$.\!\!^{\prime\prime}$}}
\def\farcm{\hbox{$.\!\!^{\prime}$}}
\def\arcmin{\hbox{$^\prime$}}
\def\arcsec{\hbox{$^{\prime\prime}$}}
\newcommand{\gsim}{\raisebox{-0.3ex}{\mbox{$\stackrel{>}{_\sim} \,$}}}
\newcommand{\lsim}{\raisebox{-0.3ex}{\mbox{$\stackrel{<}{_\sim} \,$}}}

\begin{document}

\title[Pulsar luminosities in 47 Tuc.]{The radio luminosity distribution of
pulsars in 47 Tucanae}

\author[D. McConnell et al.]{D. McConnell$^1$,
    A.A. Deshpande$^{1,2,3}$,
    T. Connors$^{4,5}$ \&
    J.G. Ables$^6$\\
$^1$ Australia Telescope National Facility, CSIRO, PO Box 76, Epping, NSW 1710,
Australia\\
$^2$ Raman Research Institute, Bangalore 560 080, India\\
$^3$ Present address: Arecibo Observatory, NAIC, HC3 Box 53995, Arecibo, Puerto Rico\\
$^4$ Department of Physics, University of Sydney, Australia\\
$^5$ Present address: Centre for Astrophysics and Supercomputing, Swinburne
University, PO Box 218, Hawthorn, Vic 3122, Australia\\
$^6$ Telecommunications and Industrial Physics, CSIRO, PO Box 76, Epping, NSW 1710,
Australia
}

\maketitle

\begin{abstract}

We have used the Australia Telescope Compact Array to seek the integrated radio
flux from all the pulsars in the core of the globular cluster 47 Tucanae. We
have detected an extended region of radio emission and have calibrated its flux
against the flux distribution of the known pulsars in the cluster. We find the
total 20-cm radio flux from the cluster's pulsars to be $S = 2.0 \pm 0.3
\mbox{mJy}$.  This implies the lower limit to the radio luminosity distribution
to be $^{min}L_{1400} = 0.4 \mbox{mJy kpc}^2$ and the size of the observable
pulsar population to be $N \lsim 30$.

\end{abstract}

\section{Introduction}

Pulsars fall into two quite distinct categories: young objects with relatively
low spin rate --- standard pulsars; and much older objects spinning several
hundred times per second --- the so-called millisecond pulsars.  The majority
of known pulsars fall into the former category and are distributed across the
Galactic disk as expected for objects that have originated fairly recently
(10$^4$ -- 10$^7$ years) from massive stellar supernovae.  The millisecond
pulsars have quite a different distribution, but which can also be understood
by looking at their progenitors. Srinivasan \& van den Heuvel (1982) and Alpar
et al. (1982) proposed that the millisecond pulsars have been spun up to high
rotation rates during an accretion phase as a member of low mass X-ray binary
(LMXB) systems.  The tendency for millisecond pulsars to be found in globular
clusters is seen as a natural consequence of the high incidence of LMXB systems
in such clusters.  Of the galactic globular clsuters, 47 Tucanae dominates as
the host of at least 22 pulsars (Manchester, 2000).  This number is large
enough  that it presents the possibility of certain population studies, some of
which have been addressed by Camilo et al. (2000) and Friere et al. (2001).

In this paper we discuss the number of potentially observable pulsars in
47 Tucanae. We use the term ``potentially observable'' (sometimes simply
``observable'') for pulsars whose emission beam passes over the Earth
and so could be detected with a sufficiently sensitive telescope. The
probability that any pulsar is potentially observable depends on the
angular width of its emission beam, and the fraction of all pulsars that
are observable relates to some population average (latitudinal) beam width. 

Camilo et al. have estimated the total number of potentially observable
pulsars in 47 Tuc to be $\sim$200.  They base their estimate on the
supposition that the cluster pulsars are distributed in luminosity
according to the law observed for the standard pulsars in the Galactic disk.
That is, their differential luminosity distribution is assumed to be
consistent with
\begin{equation} d \log N = - d  \log L  \end{equation}
where $N$ is the number of pulsars and $L$ their luminosity. Camilo et al. also
assumed that the pulsars extend in luminosity down to the lower limit
observed in the Galactic pulsar population, measured at 400~MHz, of $
^{min}L_{400} \sim 1 \mbox{~mJy kpc}^2$.

The determination of the luminosity distribution for a population of pulsars is
a difficult process (see for example Lyne, Manchester \& Taylor, 1985) for the
following reasons.  Firstly, the distance to each pulsar in the sample must be
known. This is usually inferred from the observed dispersion of the pulsar
signal and estimates of the free electron density along the interstellar line
of sight and is often given with large uncertainty.  Secondly, any sample is
incomplete because of observational selection effects.  This is particularly
relevant for the fainter pulsars in the population and introduces extra
uncertainty in the measured luminosity function at the low-luminosity end of
the distribution.

A measurement of the total radio flux from the core region of 47 Tucanae would,
if the form of the pulsar flux distribution were known, provide an estimate of
the lower limit to the distribution and of the number of pulsars in the
cluster. Specifically, if the cluster contains the estimated 200 observable
pulsars and if the flux distribution as assumed by Camilo et al. is followed,
the total radio flux from the pulsars in the cluster core should be $\sim$4~mJy
--- double that of those already detected. Here we report observations designed
to be sensitive to such a distribution of 20~cm emission from the core. The
observations reported here were made with the Australia Telescope Compact Array
(ATCA) in low resolution (short baseline) configurations and are complementary
to earlier higher resolution ATCA observations reported by McConnell \& Ables,
2000 (referred to hereafter as MA2000) and McConnell, Deacon \& Ables, 2001
(MDA2001).

\section{The pulsar luminosity distribution}
\label{distribution}

The observed radio luminosity of a pulsar is \begin{equation} L = s D^2
\end{equation} where $s$ is its radio flux and $D$ its distance from the
observer. $L$ is usually expressed in the units mJy~kpc$^2$. The
distance to 47 Tuc is $4.5\pm0.3$~kpc (Zoccali et al., 2001). In equation (2), the
distance $D$ to all pulsars in 47 Tuc is effectively the same. Given the
angular spread on the sky of the 47 Tuc pulsars $\Delta\theta \leq
2\farcm5$, and assuming the distribution to be symmetrical, the spread
in pulsar distance $\Delta D \leq D \Delta\theta \simeq 3 \mbox{pc}$ is
negligible. In the following we describe the distribution of radio
fluxes, and in conclusion give the equivalent parameters of the
luminosity distribution for comparison with the general pulsar
population. 

As noted in the introduction, the population of standard pulsars in the
Galactic disc has a luminosity distribution consistent with equation 1.
It is useful to write explicitly the underlying frequency distribution
and to determine whether the 47 Tuc pulsars with known fluxes also
belong to a population consistent with that relation.
We write the frequency distribution of pulsars, $n(s)$, with radio flux $s$ as
\begin{equation}
 n(s) ds = \left\{ \begin{array}{ll}
2Ns_{min}s^{-2}ds & s \geq s_{min} \\
0 & s < s_{min}
\end{array}
\right.
  \end{equation}
where
\begin{equation}  \int_{s_{min}}^{\infty} n(s) ds = N  \end{equation}
is the total size of the pulsar population. The cumulative frequency
distribution $N(>s)$, ie. the number of pulsars with flux density greater than
$s$, is then
\begin{equation}
N(>s) = \int_s^{\infty} n(x) dx = \frac{Ns_{min}}{s}  \end{equation}
for $s > s_{min}$. After taking the logarithm we get
\begin{equation}  \log N(>s) = \log N s_{min} - \log s  \end{equation}
which, in differential form, corresponds to equation 1. For such a
distribution, the integrated flux of all pulsars with flux above some
limit $s$ is
\begin{equation} S(>s) = N s_{min} \sum_{k=1}^{m} \frac{1}{k} \end{equation}
where $m = N s_{min}/s$. 

To verify that equation 6 describes the sample of pulsars already
detected in 47 Tuc, we have analysed their fluxes reported by Camilo et
al. (2000). We give their cumulative flux distribution $N(>s)$ in Figure
1, which shows that pulsars with $s \gsim 0.07~\mbox{mJy}$ follow the
assumed law. Fainter pulsars appear to be less numerous, as might be
expected either because of approaching the sensitivity limit of the
detection equipment or because the intrinsic lower flux limit is being
approached. We have fitted the cumulative distribution for $s \geq
0.07~\mbox{mJy}$ (9 pulsars) to an expression of the form $ \log N(>s) =
a + b \log s $ and find that $a = -0.1\pm0.2$, $b = -0.9\pm0.2$,
consistent with equation 6. Constraining the distribution to follow
equation 6 exactly ($b = -1$) we find
\begin{equation} \log N(>s) = -0.18\pm0.07 - \log s \end{equation}
This indicates that the quantity
$Ns_{min} \simeq 0.66\pm0.11 \mbox{mJy}$.

\begin{figure}
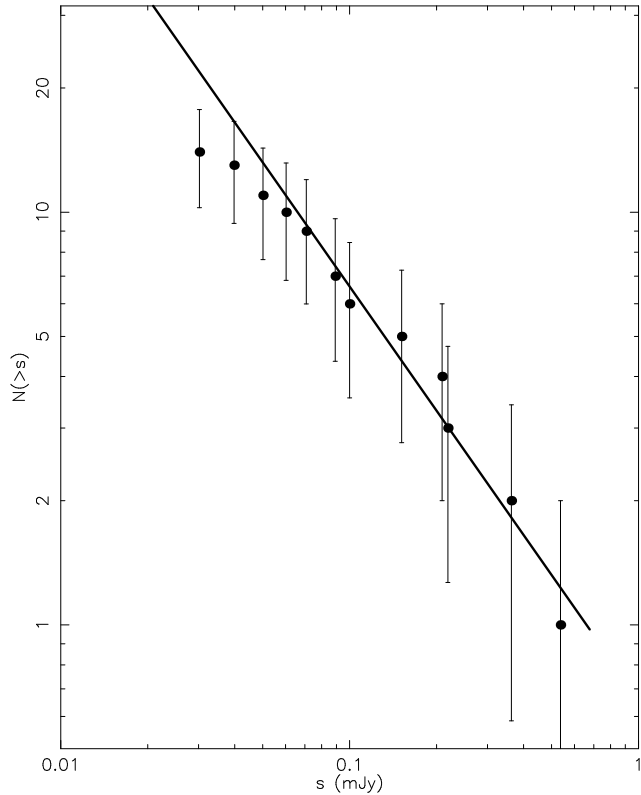

\begin{center}
\putfig{fig1}{8.4cm}
\caption{The cumulative distibution of 47 Tuc pulsars with respect to 20~cm
radio flux taken from Camilo et al. (2000). The line is a fit to the
distribution for pulsars with $s \geq 0.07 \mbox{mJy}$, constrained to have
slope $b = -1$: $\log N(>s) = -0.18 - \log s$.} \label{freqdist}
\end{center}
\end{figure}

\section{Observations and Data Analysis}

The aim of the observations reported here was to detect radio emission from the
core of 47 Tucanae.  The likely distribution of pulsars in the cluster has
already been indicated by the positions of the brighter members as determined
from pulse timing analysis (Freire et al., 2001). We sought to detect the
integrated emission of a possibly larger population of pulsars sharing this
spatial distribution.  All pulsars lie within 3.2 $r_{c}$ of the cluster
centre, where the core radius $r_{c} = 23\farcs1\pm1\farcs7$\ (Howell,
Guhathakurta \& Gilliland, 2000). We therefore chose to use the Australia
Telescope Compact Array in the 0.75 and 0.375 configurations  to give
sensitivity to brightness distributions of order $\sim$50\arcsec\ in size.  
The observations were made over a bandwidth of 128~MHz centred at 1.408~GHz
($\lambda \sim$ 20~cm).

The {\sc miriad} data reduction package was used to calibrate the data and
form images.  The antenna gains were calibrated using brief observations of
the source B2353-696 ($S_{1.4} = 1.05$Jy). The flux scale was referred to the
ATCA primary flux calibrator B1934-638 ($S_{1.4} = 14.9$ Jy). Some improvement
to the antenna gain calibration was achieved by self calibration of the final
image.

\begin{table}
\begin{center}
\caption{Summary of the ATCA observations of 47 Tucanae.}

\begin{tabular}{cccl}
\hline
\multicolumn{1}{c}{Date} & \multicolumn{1}{c}{Time} &
 \multicolumn{1}{c}{Integration} &
 \multicolumn{1}{c}{Array} \\
\multicolumn{1}{c}{(UT)} & \multicolumn{1}{c}{(UT)} &
 \multicolumn{1}{c}{(hours)} & \\
\hline
2000 Aug 05 & 14:16 & 9.0 & 0.375 \\
2000 Dec 24 & 00:10 & 11.2 & 0.75C \\
2000 Dec 25 & 23:13 & 18.7 & 0.75C \\
2000 Dec 27 & 11:59 & 8.9 & 0.75C \\
\hline
\end{tabular}
\end{center}
\end{table}

The images we present here were derived from the observations listed in Table
1. Care was taken to image a large area (137\arcmin $\times$ 137\arcmin)  to
cover the annular region of sky falling in the first sidelobe of the 22m
antenna beam pattern which lies about 60\arcmin\ from the centre of the primary
lobe.  A number of sources with flux density of up to $\sim$6~mJy were detected
in the first sidelobe.

\section{Image Analysis}

The image is dominated by the many background sources in the field, the
brightest having a flux of $\sim$50~mJy and lying 6\farcm5 from the image
centre. The effect of these sources (including their possible variability),
combined with unmodelled antenna gain variations, is to increase the image
noise from the 20$\mu$Jy/beam expected from thermal (receiver and sky)
fluctuations to the measured value of $\sigma = 60\mu$Jy/beam.  Nonetheless, a
diffuse area of 20-cm emission is visible in the cluster core region of the
image shown in Figure 2.

\begin{figure}
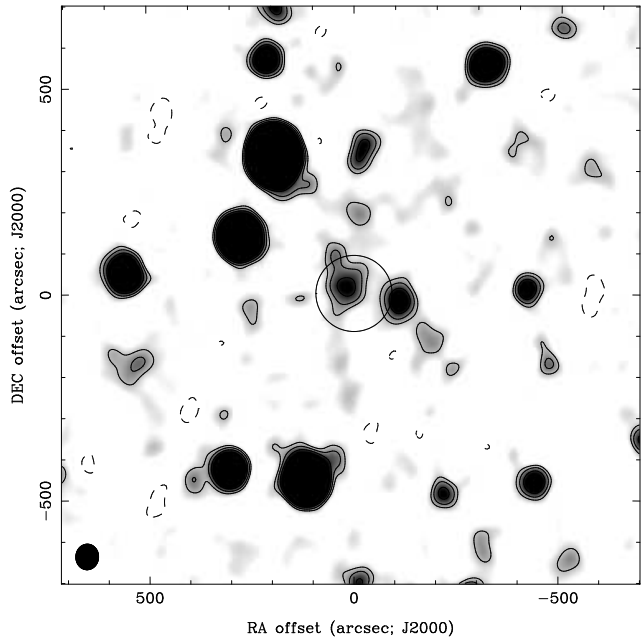

\begin{center}
\putfig{fig2}{8.4cm}
\caption{20-cm image of 47 Tucanae. The circle is centered on the cluster
centre and has  radius $4 \times r_{c}$  where
$r_{c} = 23\farcs1$. The size of the restoring beam is shown at lower left.
Image brightness is shown as contours at
-3,3,5 \& 8 $\times$ the rms fluctuation in the image of $\sigma =
60\mu\mbox{Jy}$.
}
\label{image1}
\end{center}
\end{figure}

Before measuring the flux of that emission and attributing it to the cluster
pulsars, we must consider a number of effects associated with flux calibration,
the sensitivity of the array to extended emission and pulsar variability, all
of which may influence the measurement we wish to make, and its
interpretation.

\subsection{Sensitivity to extended emission}

The expected spatial extent of pulsar radio emission is indicated by the
radial distribution of the 15 pulsars with locations derived from pulse
timing analysis (Friere et al, 2001).  These pulsars, to an approximation
sufficient for our purposes, are uniformly distributed in radial distance
from the cluster centre out to 3.2$r_{c}$.  Thus half the pulsar radio flux
is expected from the inner quarter of a circle of radius $\sim75\arcsec$.
To test the sensitivity of the ATCA to the pulsar flux we constructed a
simulated visibility dataset using the same antenna locations, observation
hour angles and receiver noise characteristics as those of the real
observations.  To these data we added contributions corresponding to: (i) a
circular source with total flux 1~mJy and a gaussian profile of full width
to half maximum of 75\arcsec; (ii) a set of 17 point sources whose fluxes
(each $<$100~$\mu$Jy, total 946~$\mu$Jy) were drawn from the distribution
described in section 2 and whose positions were drawn from the uniform radial
distribution mentioned above.  In the images derived from both these
trial datasets the measured flux of the added component differed from the
expected value by less than the standard error expected from the simulated
thermal noise.  We conclude that the baseline lengths used in the observations
adequately sample the spatial scales present in the likely pulsar emission.

\subsection{Flux calibration}
\label{calibration}

We aim to compare the radio flux of pulsars measured with the Parkes
radiotelescope (Camilo et al., 2000) against the flux of radio sources in 
images formed with the ATCA.  Therefore it is necessary to ensure consistency
between the flux scales used in the two sets of measurements. In the case of
the Parkes observations the pulsar time series signal strength is compared to
the amplitude of noise fluctuations arising from components of the receiving
equipment and from the sky --- the so-called system equivalent flux density. 
Camilo et al. estimate a systematic uncertainty of order 25\% in their
published flux scale. The amplitudes of the visibilities measured with the ATCA
were calibrated using observations of the source B1934-638. No specific attempt
has been made to tie together the flux scales of the two instruments (F.
Camilo, private communication).

To make a direct comparison we use flux measurements of the five strongest
pulsars (C, D, E, F and J) with both instruments: the Parkes measurements are
those of Camilo et al. (2001); the ATCA measurements are from the earlier high
resolution observations (MDA2001) which, unlike the observations described
here, used the 6-km configurations of the ATCA to achieve its highest possible
resolution.  The measurement results appear in Table 2. We use the symbol $S$
for integrated fluxes, distinguishing between the bright (five brightest) and
faint (sixth brightest and fainter) pulsars with the leading superscripts $b$
and $f$; hence $^{b}S_P$ is the sum of fluxes for pulsars C, D, E, F and J
measured at Parkes, and $^{f}S_A$ is the ATCA measurement of flux from all
fainter pulsars. We calculate the total fluxes: \begin{equation} ^{b}S_{P} =
1.48 \pm 0.09 \mbox{mJy}  \end{equation} \begin{equation} ^{b}S_{A} = 1.20 \pm
0.16 \mbox{mJy}  \end{equation} To validly compare fluxes measured in the ATCA
images with the flux distribution quoted by Camilo et al. we scale the ATCA
data by the factor $^{b}S_{P}/^{b}S_{A} = 1.23 \pm 0.18$.  This factor is
consistent with the Parkes flux scale uncertainty of $\sim$25\% estimated by
Camilo et al.

\subsection{Pulsar variability}
\label{variability}

A remarkable feature of the 47 Tuc pulsars is their strong scintillation.
Camilo et al. (2000) report that pulsar fluxes measured over a series of
17.5-minute observations may vary by factors of up to $\sim$30. In MA2000 we
reported variablilty of the imaged pulsars over a series of 12-hour
integrations. We have extended that analysis using the additional ATCA
observations described in MDA2001 and find that the measured fluxes of an
individual pulsar from a set of $\sim$12-hour integrations has a standard
deviation approximately equal to the mean flux of that pulsar. Note that for
diffractive scintillation, the expected exponential distribution of fluxes
(Rickett, 1977) has standard deviation equal to the mean, consistent with this
observation. Thus, even over the 48-hour integration of Figure 1, scintillation
of the brightest pulsars in the cluster could significantly influence the
integrated flux from the core and take us to a misleading conclusion. 

It is relevant to consider the variability of total flux of the cluster's
pulsar population.  Consider a set of variable sources each with mean flux
$s_i$ and rms fluctuations in flux of $\delta s_i \sim f \times s_i$ after
integration over time $\tau$, where $f$ is fixed for the whole population. 
Then if the sources have a luminosity function as given by equation (1), it
follows that the rms fluctuation of the total flux is 

\begin{equation}
\Delta S \simeq 1.3 f s_{max}
\end{equation}

where $s_{max}$ is the flux of the brightest member of the set.  In the case of
the 47 Tuc pulsars $f \sim 1$ for 12-hour integrations, so we would expect
12-hour measurements of the total flux to fluctuate with $\Delta S \sim 0.7
\mbox{mJy}$.  This quantity is dominated by scintillations of the brightest
pulsars in the cluster.  Independent measurements of their individual fluxes
over the integration period considered would improve the estimate of the mean
total flux, provided the measurement error is less than the expected rms
fluctuation over the same period.

The image presented in Figure 1 was observed with five elements of the ATCA
arranged to give baselines of up to 750m, chosen to give best brightness
sensitivity to extended emission, but which naturally lacks the angular
resolution to measure fluxes of individual pulsars. However, the ATCA has a
sixth element fixed permanently 3km from the rail track supporting the moveable
five, so that each observation included visibility measurements
from five baselines with lengths in the ranges 4.3 -- 5.0 km (0.75C array) or
5.5 -- 6.0 km (0.375 array). The image derived from all baselines, inspite of
the poor sampling of the uv-plane, has a synthesised beam of
4\farcs1$\times$3\farcs5 and so is sensitive to point sources. The image has
noise fluctuations of $\sigma = 40 \mu$Jy/beam and all of the brightest five
pulsars are visible at $>3 \sigma$ (Table 2, column 4). Thus flux measurement
of the five brighter pulsars introduces significantly less uncertainty ($\sim
\sqrt{5} \times 0.04$~mJy) than expected from the scintillations. 

\subsection{Image quality in the presence of variable sources}
\label{field}

The fidelity of radio aperture synthesis images depends on the constancy of the
region being imaged.  In reality variable sources exist and some degradation of
any 20cm image from the ATCA can be expected.  For these observations we expect
the pulsars themselves to vary.  The diffraction responses around a variable
source will differ from the point spread function of the observation and so
will not be modelled correctly during deconvolution.  In aperture synthesis the
point spread function has zero mean, so that the effect on the image some
distance from the varying source will be as often positive as negative.  The
effect of a number of independently varying sources in the field is to increase
the background fluctuations in the image.  The images presented here have rms
brightness fluctuations $\sigma = 60 \mu$Jy/beam, compared with the value
$\sigma = 20 \mu$Jy/beam expected from thermal considerations alone.  We can
attribute at least part of this extra noise to variations in the background
sources.

To gauge the size of fluctuations induced by source variability we have
artificially added a 0.5mJy source to the observed data over a 11-hour section
of the 48-hour integration and determined the effect  on the flux measured in
the core of the cluster.  With the artificial source placed in the core, the
measured flux is increased by the expected time averaged flux of
$\sim$0.12~mJy. If the test source is placed outside the core region, the
influence is position dependent, but the change to the measured core flux is
$\lsim 25\mu$Jy.

\begin{table}
\begin{center}
 \caption{Flux measurements of the pulsars in 47 Tucanae. (2): fluxes
measured at Parkes (Camilo et al., 2001). (3):
fluxes measured from the ATCA image reported in MDA2001. (4):
fluxes in the image formed from the long baseline data as part of the
observations listed in Table 1.  Uncertainties in the last digits quoted
in (2),(3) appear in parentheses.  The uncertainties of values in (4) are
$\sim 0.04$~mJy.}

\begin{tabular}{clll}
\hline
\multicolumn{1}{c}{Pulsar} &
\multicolumn{1}{c}{$S_{Pks}$} &
\multicolumn{1}{c}{$S_{ATCA}$} &
\multicolumn{1}{c}{$S\prime_{ATCA}$} \\

\multicolumn{1}{c}{name} &
\multicolumn{1}{c}{(mJy)} &
\multicolumn{1}{c}{(mJy)} &
\multicolumn{1}{c}{(mJy)} \\

\multicolumn{1}{c}{(1)} & \multicolumn{1}{c}{(2)}
 & \multicolumn{1}{c}{(3)} & \multicolumn{1}{c}{(4)} \\
\hline
C & 0.36(4) & 0.24(6)  & 0.13 \\
D & 0.22(3) & 0.37(10) & 0.22 \\
E & 0.21(3) & 0.08(2)  & 0.14 \\
F & 0.15(2) & 0.18(5)  & 0.14 \\
G & 0.05(2) & &  \\
H & 0.09(2) & &  \\
I & 0.09(1) & &  \\
J & 0.54(6) & 0.33(9) & 0.16 \\
L & 0.04(1) & &  \\
M & 0.07(2) & &  \\
N & 0.03(1) & &  \\
O & 0.10(1) & &  \\
Q & 0.05(2) & &  \\
U & 0.06(1) & &  \\
\hline
\end{tabular}
\end{center}
\end{table}

\subsection{Analysis method}
\label{method}

To determine the best possible estimate of the total pulsar flux, on a
scale that allows direct comparison with the flux distribution analysed
in section \ref{distribution}, we have analysed the data in the steps
described below.
\begin{itemize}
\item Measure the fluxes for all pulsars above a $3\sigma$
detection limit in the high resolution image described in section
\ref{variability}.
\item Subtract from the low resolution visibility data, a set of point
source responses with the measured fluxes at the corresponding pulsar
locations.
\item Subtract from the low resolution visibility data all source components
detected in the high resolution image outside the circular region centred on the
core with radius $4r_c$.
\item Subtract an additional source (S176 in MDA2001) which lies within $4r_c$
of the cluster centre.  This source was discussed in MDA2001.  It lacks
variability, has a flat spectrum and is not suspected of being a pulsar.
\item Derive the low resolution image from the modified visibility
dataset, and measure the flux of all emission within $4r_c$ of the cluster
centre.
\item Correct this figure for the flux scale diferences between the ATCA
and Parkes measurements (see section \ref{calibration}).
\item Add the mean flux,as reported by Camilo et al. (2000), of each
pulsar detected and subtracted in the first two analysis steps above.
\end{itemize}

\section{Results}

Figure 3  shows the central part of 47 Tucanae at 20~cm before (left) and after
(right) the pulsar and source removal described in section \ref{method}. 
Significant emission is visible from only a small part of the $4r_{c}$ circle
around the cluster centre and the total radio flux is much less than the
$\sim$4~mJy expected if the predicted $\sim$200 pulsars exist in the cluster. 
The total flux measured in the residual image inside the $4r_c$ circle is 
$^{f}S_A = 0.42 \pm 0.22 \mbox{mJy}$.  The uncertainty is calculated from the
observed brightness noise in the image (0.06~mJy/beam), the error in
determination of the bright pulsar flux, and an estimate of the likely
difference in flux of a 48-hour integration from the long term mean flux of the
weaker, unsubtracted pulsars (see section \ref{variability} and equation
(11)).  Converting the measurement to the Parkes flux scale using the factor
derived in section \ref{calibration}, we have  $^{f}S_P = 0.52 \pm 0.28
\mbox{mJy}$.  Thus the total 20-cm radio flux is

\begin{equation}
 S_P = ^{b}S_P + ^{f}S_P = 2.00 \pm 0.29 \mbox{mJy}
\end{equation}

\begin{figure}
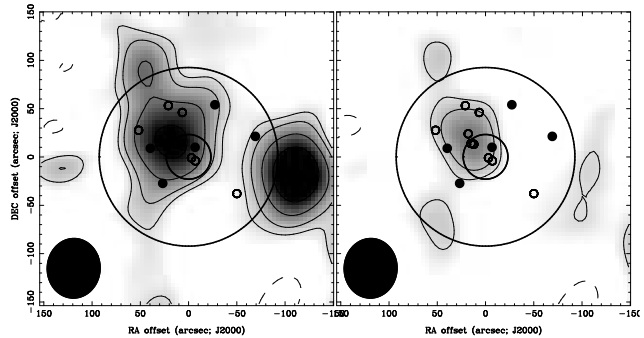

\begin{center}
\putfig{fig3}{8.4cm}

\caption{Image of the core region of 47 Tucanae. The left panel shows the radio
emission from all pulsars and background sources.  On the right is the residual
image after subtraction of components outside the $4r_c$ circle, pulsars
C,D,E,F and J, and source S176 (see text).  Image coordinates are indicated as
offsets from the cluster centre: $\alpha_{J2000} = 00:24:05.9$, $\delta_{J2000}
= -72:04:51.1$.  Image brightness is shown as contours at -2,2,3,5 \& 8
$\times$ the rms fluctuation in the image of $\sigma = 60\mu\mbox{Jy}$. The
cluster core $r_{c} = 23\farcs1$ and an area of radius $4 \times r_{c}$ are
indicated.  The positions of the five brightest pulsars (C, D, E, F, J) are
marked with $\bullet$. All other known pulsar positions are marked $\circ$. The
size of the restoring beam is shown at lower left.}

\label{image2}
\end{center}
\end{figure}

In Figure 4 we relate this result to the cumulative flux distribution.  The
vertical line indicates the likely lower bound to the flux of pulsars in the
cluster.  It is unlikely that the flux distribution extends below $s_{min}
\simeq 0.02$~mJy, corresponding to an upper bound for the pulsar population
size of $N \lsim 30$.  This result is consistent with the possibility that
there are no further radio pulsars to be detected in 47 Tuc.

The minimum luminosity of this distribution is
$^{min}L_{1400} = s_{min} D^2 = 0.4 \mbox {mJy kpc}^2$ at 1400~MHz.  Pulsar
luminosities are often referred to 400~MHz with a spectral
index of $-2$ assumed to allow conversion:
$L_{400} = L_{1400} (400/1400)^{-2}$.  Using this approach we estimate the
lower limit of 400~MHz luminosity of the 47 Tuc pulsar population to be
$^{min}L_{400} = 6 \mbox{ mJy kpc}^2$.

\begin{figure}
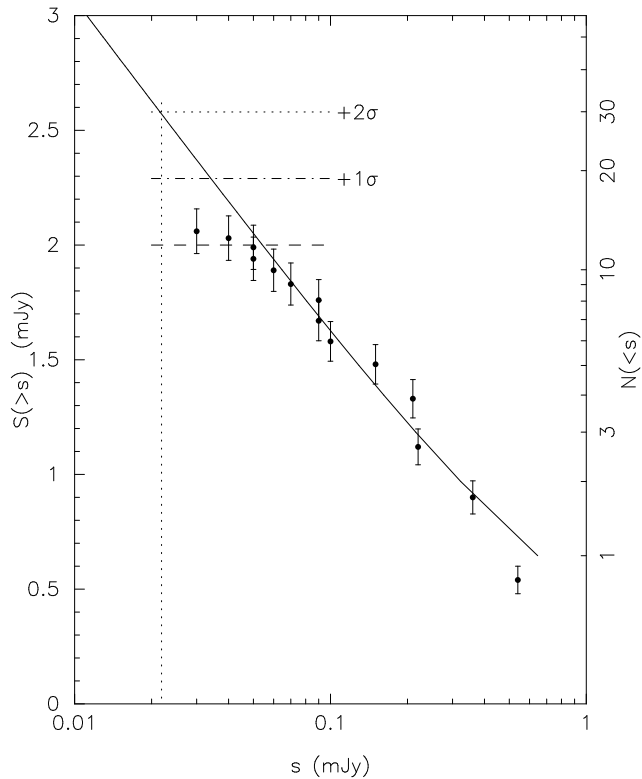

\begin{center}
\putfig{fig4}{8.4cm}

\caption{The cumulative flux distribution $S(>s)$ of pulsars in 47 Tuc
as measured by Camilo et al., 2000. The solid line shows the expected
value of $S(>s)$ corresponding to expression 7 using the fitted value of
the product $N s_{min} = 0.66 \mbox{mJy}$ from Figure 1. Corresponding
values of the total pulsar number are shown on the right hand axis. The
best estimate of total radio flux from the cluster core is indicated by
the dashed line. The horizontal dotted line shows the $2\sigma$ upper
bound. The vertical dotted line indicates the likely lower bound to the
pulsar flux distribution.}
\label{totalflux}
\end{center}
\end{figure}

\section{Discussion}

In this work we have estimated two parameters of the population of
pulsars in 47 Tucanae: the lower bound to its luminosity distribution
$^{min}L_{1400}$ and its size $N$, the total number of observable
pulsars in the cluster.  Our technique is similar to that used by
Fruchter and Goss (1990) who made measurements of integrated
flux densities of a number of globular clusters and estimated the number
of pulsars in each cluster.  Fruchter and Goss used a luminosity
function corresponding to equation 3 with an assumed value of
$^{min}L_{1400} \sim 0.2 \mbox{ mJy kpc}^2$, the luminosity of the
weakest millisecond pulsar known at the time of their work.  In the case
of 47 Tuc, we have used our measurement of the integrated flux and also
the measured fluxes of the brightest members of the population to
constrain both N and $^{min}L_{1400}$. We find
$^{min}L_{1400} \sim 0.4 \mbox{ mJy kpc}^2$, significantly greater than
the value assumed by Fruchter and Goss.

As we noted in the introduction, the determination of the radio luminosity
distribution of pulsars is frustrated by uncertainties in pulsar distances and
the difficulty of defining a sample free of strong selection effects.  To make
this determination usefully for all currently known millisecond pulsars is
probably impossible.  The pulsars in 47 Tuc provide the best opportunity
available at present to characterise the radio luminosities of millisecond
pulsars, and perhaps of pulsars in general.

If we accept the inferences made about the luminosity function of the 47 Tuc
pulsar population, then the least luminous members of the population are nearly
an order of magnitude brighter than the weakest pulsars in the general galactic
population.  We speculate that the origin of this difference relates more to
the distinction between standard and millisecond pulsars, than to the different
locations of the sample populations.  The radio luminosity of an individual
pulsar is expected to decline steadily, in accord with declining rotational
energy.  Ultimately the rotational energy reduces to the point of being
insufficient to power the radio emission mechanism, and the pulsar crosses the
``death line'' --- see for example the discussion by Chen and Ruderman (1993). 
As a class, the millisecond pulsars have very small values of $\dot{P}$, so
that their evolution towards the death line is very slow.  The characteristic
age at which a pulsar crosses the death-line (which also corresponds to a fixed
mechanical luminosity line), can be shown to be inversely proportional to its
magnetic field strength.  Hence, compared to normal pulsars the age at which
radio millisecond pulsars die would be several orders of magnitude
longer.  Perhaps the relatively high low-luminosity bound inferred for the 47
Tuc pulsars indicates how much additional time is required for them to evolve
through lower luminosities to the death line.

Most energy lost from a pulsar is radiated in the form of low frequency
``dipole radiation'', generated by the rotation of the star's magnetic field
and observable as the gradual increase in rotation period.  A tiny fraction of
the energy loss is visible in the radio spectrum.  Real physical insight might
be expected from successfully connecting, through measurement, the radio
luminosity with the total energy loss rate.  The pulsars in 47 Tuc present us
with a measurable radio luminosity distribution at 1.4~GHz.  Grindlay et al.
(2002) estimated energy loss rates $\dot{E}$ of the pulsars after assuming that
the location of each pulsar in the cluster's gravitational potential can be
estimated from its differential dispersion measure.  If the total radio
luminosities could be inferred from the 1.4~GHz measurements and the likely beam
solid angles, some attempt could be made to connect them to the estimates of
$\dot{E}$.

Finally we note that X-ray detection provides an alternate means of estimating
the total population of 47 Tuc.  Grindlay et al. (2002) have reported
\emph{Chandra} observations and place an upper limit of 90 pulsars in the
cluster.  The X-rays are attributed to thermal emission from the polar cap and
are not beamed.  Adopting the upper limits of 90 X-ray and 30 radio pulsars, and
assuming the X-rays are not beamed, the implied radio beaming factor is
$\sim0.3$.

In conclusion, we have determined a lower bound to the luminosity of pulsars in
the globular cluster 47 Tucanae.  If this bound is typical of millisecond
pulsars in general, then it would imply a considerably smaller number of
pulsars in globular clusters than believed earlier.

\section*{Acknowledgements}

We thank the staff of the ATCA for their expert support during this
observational programme. DM is grateful for the support of the RRI during his
one month study visit, during which much of this paper was written. AAD is
grateful to the ATNF for support during the study visit to Australia, during
which this project was initiated. We thank the referee for suggestions which
led to improvements in this paper. TC was supported by an ATNF Summer Vacation
scholarship. The Australia Telescope Compact Array is funded by the
Commonwealth of Australia for operation as a National Facility by CSIRO.

\bibliographystyle{mnras}

\end{document}